\documentclass[10pt]{iopart}
\usepackage{iopams}
\usepackage{graphicx}
\usepackage{cite}
\usepackage[pdfstartview=FitH, pdfpagemode=None, pdfpagelayout=OneColumn, colorlinks=true, linkcolor=blue, citecolor = blue]{hyperref}
\newcommand{\text}[1]{\mathrm{#1}}
\newcommand{\eqref}[1]{(\ref{#1})}
\def\ds{\displaystyle}

\begin{document}

\title[Electron-seeded self-modulation in ramped-density plasma]{Proton beam self-modulation seeded by electron bunch in plasma with density ramp}

\author{K.V. Lotov}
\address{Budker Institute of Nuclear Physics, Novosibirsk, 630090, Russia}
\address{Novosibirsk State University, Novosibirsk, 630090, Russia}
\ead{K.V.Lotov@inp.nsk.su}
\author{V.A. Minakov}
\address{Budker Institute of Nuclear Physics, Novosibirsk, 630090, Russia}
\address{Novosibirsk State University, Novosibirsk, 630090, Russia}
\ead{V.A.Minakov@inp.nsk.su}

\vspace{10pt}
\begin{indented}
\item[]\today
\end{indented}

\begin{abstract}
Seeded self-modulation in a plasma can transform a long proton beam into a train of micro-bunches that can excite a strong wakefield over long distances, but this needs the plasma to have a certain density profile with a short-scale ramp up. For the parameters of the AWAKE experiment at CERN, we numerically study which density profiles are optimal if the self-modulation is seeded by a short electron bunch. With the optimal profiles, it is possible to ``freeze'' the wakefield at approximately half the wavebreaking level. High-energy electron bunches (160\,MeV) are less efficient seeds than low-energy ones (18\,MeV), because the wakefield of the former lasts longer than necessary for efficient seeding.
\end{abstract}

\vspace{2pc}

\ioptwocol
\section{Introduction}
\label{s1}

Plasmas can withstand orders of magnitude stronger electric fields than those attainable in solid accelerating structures \cite{Phys.Today-56(6)-47,Phys.Today-62(3)-44,RMP81-1229,RMP90-035002}. This feature opens up prospects for either reducing the size of linear particle accelerators or increasing the energy of lepton beams to the level currently achieved by proton synchrotrons, or both. Plasma-based accelerators differ in type of driver, that is, an object that creates the strong, correctly phased field in the plasma, called a wakefield. Each type of driver has its own advantages \cite{NatPhot7-775,RAST9-63}, and the advantage of proton beams is their large energy, which makes it possible to accelerate particles to TeVs in a single plasma cell, without staging of many plasma sections to reach the desired energy \cite{NatPhys5-363,PPCF56-084013,RAST9-85}.

The proton driven plasma wakefield acceleration is experimentally studied at the AWAKE facility at CERN \cite{NIMA-829-3, NIMA-829-76, PPCF60-014046}. The first experiments showed the possibility of exciting the wakefield using the proton beam of the Super Proton Synchrotron (SPS) \cite{PRL122-054801,PRL122-054802} and accelerating electrons up to 2\,GeV in a plasma cell 10\,m long \cite{Nat.561-363,PTRSA377-20180418}. The SPS proton bunches are hundreds of times longer than the plasma wavelength and efficiently drive the wave only after they are micro-bunched as a result of seeded self-modulation (SSM) \cite{EPAC98-806,PRL104-255003,PRL107-145002,PRL107-145003,PoP22-103110,PPCF56-084014,PoP20-103111}. In a longitudinally uniform plasma, the micro-bunches are formed in such a way that they efficiently drive the wave, but partly fall into its defocusing phases and, therefore, cannot propagate over a long distance \cite{PoP18-024501,PoP18-103101}. However, using a ramped or stepped up \cite{PoP18-024501} plasma density profile, it is possible to transform the proton bunch into a long-lived micro-bunch train in which all micro-bunches are focused by the wave. This train can excite the wakefield of constant amplitude over a long distance \cite{PoP18-103101}, as is to be demonstrated in future AWAKE experiments.

In the past AWAKE experiments, the SSM was initiated (seeded) by a short laser pulse that created a plasma ionization front co-propagating with the proton bunch \cite{NIMA-740-197,PPCF60-014046}. The wave phase and micro-bunch positions were locked to and controlled by the laser pulse \cite{PRL122-054802}. However, the propagation distance of the ionization front is limited by pulse depletion and cannot scale to much beyond 10~meters \cite{PRA99-063423}. Thus, for future AWAKE experiments and possible applications of proton wakefield acceleration, other methods of plasma creation should be chosen \cite{RAST9-85,PPCF60-075005} in combination with other seeding methods. One of discussed options is SSM seeding with a short electron bunch \cite{PRST-AB16-041301,NIMA-909-67,arXiv:2002.02189,arXiv:1911.07534}.

So far, all studies of SSM control using density profiles were limited to configurations in which the source of seed perturbation is located near the maximum of proton beam current \cite{PoP18-103101,PoP20-103111,PoP22-103110,PoP22-123107,JPCS1067-042009}. In this paper, we numerically study whether it is possible and how to control the SSM seeded by a short electron bunch, which density profiles are required, and what new physical effects appear in this system. The  electron seeding differs from the earlier considered cases in that the electron bunch must propagate ahead of the proton beam. Otherwise, uncontrolled self-modulation of the proton beam head can destroy the rest of the beam \cite{NIMA-410-461,PPCF56-084014}.

The paper is organized as follows. In section~\ref{s2}, we describe how we simulate the SSM and optimize the plasma density profiles. In section~\ref{s3}, we characterize the found optimum cases and possible deviations from them. We observe that high-energy electron beams are worse seeds, and discuss the reason for this in section~\ref{s4}. In section~\ref{s5}, we summarize the main findings.

\begin{figure}[tb]
\centering\includegraphics{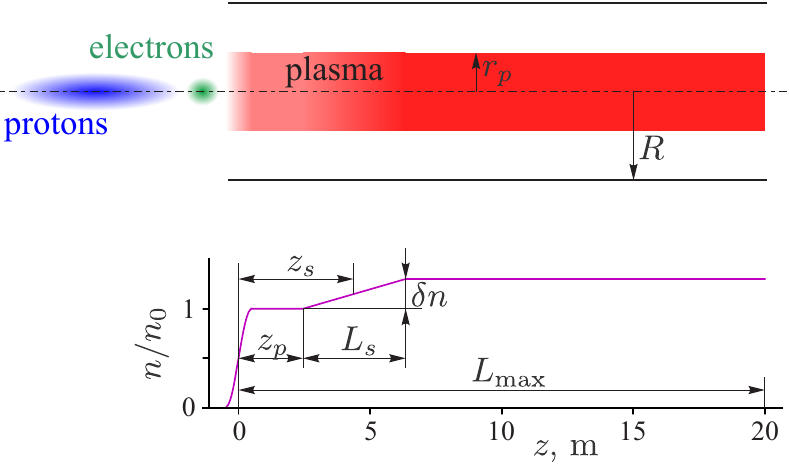}
\caption{Geometry of the problem.}\label{fig1-geometry}
\end{figure}

\section{Methods}
\label{s2}

We consider two beams, electron and proton, propagating through a long plasma cell in the $z$-direction (figure~\ref{fig1-geometry}). The problem is axisymmetric, so we use the cylindrical coordinates $(r, \varphi, z)$ and the co-moving coordinate $\xi = z-ct$, where $c$ is the speed of light. The parameters of the beams and plasma are close to those considered in the context of future AWAKE experiments (table~\ref{t1}).

The plasma is initially cold and consists of single ionized rubidium atoms. It is radially uniform up to the radius $r_p$ and has a steep boundary there, as follows from the theory \cite{NIMA-740-197,PRA99-063423}. The longitudinal plasma density profile $n(z)$ is
\begin{equation}\label{e1}
    \begin{array}{rl}
        |z| \le z_p: &
            \ds \frac{n_0}{2} \left( 1 + \frac{z/D}{\sqrt{(z/D)^2+0.25}} \right), \\
        z_p < z \le z_p + L_s: &
            \ds n_0 \left(1 + \delta n \frac{z - z_p}{L_s} \right), \\
        z_p + L_s < z: & n_0 (1 + \delta n),
    \end{array}
\end{equation}
where the parameter $D$ (diameter of the inlet orifice) determines the density profile near the entrance to the plasma cell \cite{JPD51-025203}, and the parameters $z_p$, $L_s$, and $\delta n$ (figure~\ref{fig1-geometry}) are adjusted to maximize the wakefield established after self-modulation. We analyze two options for the initial plasma density $n_0$. The low density ($2 \times 10^{14}\text{cm}^{-3}$) is of interest because it allows to visualize the micro-bunches with a streak camera \cite{PRL122-054802}. The high density ($7 \times 10^{14}\text{cm}^{-3}$) gives stronger wakefields. 
Taking into account the gradual density increase near the entrance is important for correct simulations of electron beam equilibration that occurs there. We do not consider more complex density profiles \cite{JPCS1067-042009} to keep the computational cost of optimization within reasonable limits.

\begin{table}[tb]
\caption{Parameters of the beams and plasma.}\label{t1}
 \begin{center}\begin{tabular}{ll}\hline
  Parameter and notation & Value \\ \hline
  \textit{Plasma:} &\\
  Initial plasma density, $n_0$ & $7 (2) \times 10^{14}\,\text{cm}^{-3}$ \\
  Plasma length, $L_\text{max}$ & 20\,m \\
  Plasma radius, $r_p$ & 1.4\,mm \\
  Orifice diameter, $D$ & 10\,mm \\
  Plasma ion mass number & 85 \\ \hline
  \textit{Proton beam:} &\\
  Population, $N_b$ & $3\times 10^{11}$ \\
  Length, $\sigma_{zb}$ & 7\,cm \\
  Radius, $\sigma_{rb}$ & 0.2\,mm \\
  Peak density, $n_b$ & $6.8 \times 10^{12}\,\text{cm}^{-3}$ \\
  Energy, $W_b$ & 400\,GeV \\
  Energy spread, $\delta W_b$ & 0.035\,\% \\
  Normalized emittance & 2.2\,mm\,mrad \\ \hline
  \textit{Low-energy electron beam:} &\\
  Population, $N_e$ & $3.125\times 10^9$ (500\,pC) \\
  Length, $\sigma_{ze}$ & 0.66\,mm \\
  Radius, $\sigma_{re}$ & 0.25\,mm \\
  Peak density, $n_e$ & $4.8 \times 10^{12}\,\text{cm}^{-3}$ \\
  Energy, $W_e$ & 18\,MeV \\
  Energy spread & 0\,\% \\
  Normalized emittance & 4\,mm\,mrad \\ \hline
  \textit{High-energy electron beam:} &\\
  Population, $N_e$ & $0.625\times 10^9$ (100\,pC) \\
  Length, $\sigma_{ze}$ & 0.06\,mm \\
  Radius, $\sigma_{re}$ & 0.2\,mm \\
  Peak density, $n_b$ & $1.7 \times 10^{13}\,\text{cm}^{-3}$ \\
  Energy, $W_e$ & 160\,MeV \\
  Energy spread & 0\,\% \\
  Normalized emittance & 2\,mm\,mrad \\ \hline
  Distance between beams, $\xi_e$ & 17.5\,cm \\
  \hline
 \end{tabular}\end{center}
\end{table}

In future AWAKE experiments, the plasma will likely consist of separate self-modulation and acceleration cells \cite{NIMA-909-102, arXiv:1911.07534,arXiv:1912.00779}, between which the witness electron beam is injected. However, the gap between the cells must be as small as possible in order to avoid loss of field quality in the second (acceleration) cell \cite{IPAC16-2557,PoP25-093112}. Therefore, we neglect the effect of the gap on the proton beam and consider the plasma as a single unit. As we will see later, the seed electron bunch becomes depleted long before the gap located at $z \sim 10$\,m, so the effect of the gap on this bunch can also be neglected.

We assume that the proton beam is longitudinally compressed in the SPS ring \cite{IPAC13-1820,NIMA-740-48}, as shorter beams are less demanding on plasma uniformity in the acceleration section \cite{PoP20-013102}. For the seed electron bunch, we consider two options: a low-energy bunch, used as a witness in past AWAKE experiments \cite{NIMA-829-73}, and the high-energy short bunch which can be produced with X-band accelerating structures \cite{NIMA-909-102}. Both proton and electron beams are focused to the point $z=0$. The bunch shape (density distribution) at the entrance to the plasma is
\begin{eqnarray}\nonumber
 \fl n_{\alpha} (r, \xi) = \frac{\ds N_{\alpha} e^{-r^2/2 \sigma_{r\alpha}^2}}{2 (2 \pi)^{3/2} \sigma_{r\alpha}^2 \sigma_{z\alpha}} & \left[  1 - \cos \left( \sqrt{\frac{\pi}{2}} \frac{\xi}{\sigma_{z\alpha}}  \right)  \right], \\
\label{e2}
  & -2 \sigma_{z\alpha} \sqrt{2\pi} < \xi < 0,
\end{eqnarray}
where $\alpha = b, e$, and the coordinate $\xi$ is measured from the heads of the bunches.

As a measure of the wakefield strength, we take the wakefield potential $\Phi$, the gradient of which contains information on both longitudinal ($F_\parallel$) and transverse ($F_\perp$) forces acting on an axially moving ultra-relativistic elementary charge $e$:
\begin{equation}
    F_\parallel = e E_z = -e \frac{\partial \Phi}{\partial \xi}, \quad
    F_\perp = e (E_r - B_\varphi) = -e \frac{\partial \Phi}{\partial r},
\end{equation}
where $\vec{E}$ and $\vec{B}$ are the electric and magnetic fields. Being an integral of the fields, the wakefield potential is less noisy than $E_z$ \cite{NIMA-829-3}, which facilitates numerical optimization. In a periodic plasma wave of wavelength $2 \pi c / \omega_p$, the potential unit $mc^2/e$ corresponds to the wavebreaking field $E_0 = mc\omega_p/e$, where $\omega_p = \sqrt{4 \pi n_0 e^2/m}$ is the plasma frequency and $m$ is the electron mass. We denote by $\Phi_\text{loc} (z)$ the local maxima of the potential that are reached at the points $\xi = \xi_\text{loc}$ for a certain value of $z$, and by $\Phi_\text{max} (z)$ the highest of them. The value $\Phi_f = \Phi_\text{max} (L_\text{max})$ is the optimization target function, which we maximize by changing $z_p$, $L_s$, and $\delta n$.

We use two-dimensional (axisymmetric) quasistatic code LCODE \cite{PRST-AB6-061301,NIMA-829-350,lcode}, in which plasma fields are calculated as functions of $r$ and $\xi$ with some periodicity $\Delta z$ in $z$. The simulation window is approximately $5 \sigma_{zb}$ long and $30 c/\omega_p$ wide. The wide window is necessary for the correct accounting of electrons escaping from the plasma in the event of wave breaking \cite{PRL112-194801}. 
Unless stated otherwise, $\Delta z = 200 c/\omega_p$, and the grid step of the plasma solver is $0.01 c/\omega_p$ in both $r$- and $\xi$-directions Such a small grid step is required to reduce accumulation of errors in long simulation windows \cite{IPAC13-1238}, since the fields are calculated layer-by-layer starting from the beam head. In the plasma region, there are 10 radius-weighted plasma macro-particles of each sort (electrons and ions) per cell, distributed equidistantly in $r$.  Depending on the studied regime, there are $(1.4 \div 2.6) \times 10^7$ equal macro-particles in the proton beam and $(1.8 \div 18) \times 10^4$ in the electron beam.

\begin{figure}[tb]
\centering\includegraphics{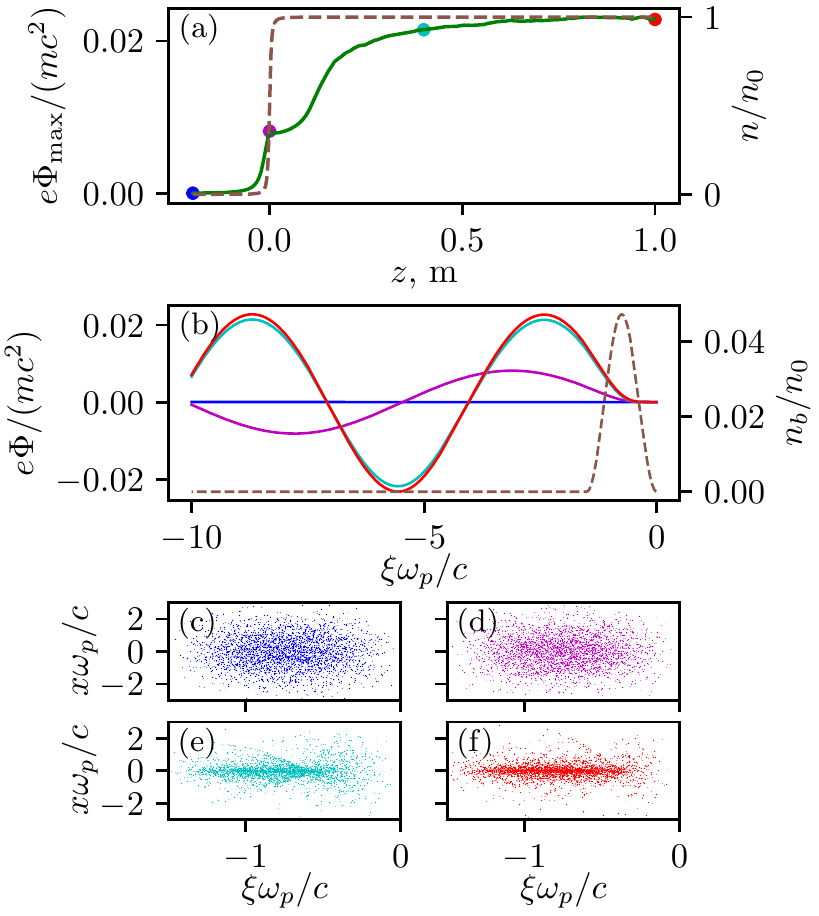}
\caption{Establishment of the radial equilibrium of $160$\,MeV electron beam in the plasma with $n_0 = 7 \times 10^{14}\text{cm}^{-3}$: (a) wakefield amplitude $\Phi_\text{max} (z)$ (solid line) and plasma density profile $n (z)$ (dashed line), (b) on-axis wakefield potential $\Phi (\xi)$ for several values of $z$ [marked in (a) by dots of the same color] and initial profile of the on-axis electron beam density $n_b(\xi)$ (dashed line); beam portraits at the considered points: $z=-1000c/\omega_p = -0.2$\,m (c),  $z=0$ (d),  $z=2000c/\omega_p = 0.4$\,m (e),  $z=5000c/\omega_p = 1$\,m (f).}\label{fig2-initial}
\end{figure}

Quasistatic codes are very efficient if the timescale of beam evolution is much longer than the plasma period $2 \pi/\omega_p$, which is the case for the proton beam. However, the presence of the seed electron beam can reduce the efficiency, since low-energy electrons must be propagated with a short time step. We avoid this problem with two tricks. First, we simulate the initial evolution of the electron beam separately in a short window ($\sim 5 \sigma_{ze}$) with a short step $\Delta z$. Second, we use beam substepping, that is, we propagate beam particles with energy-dependent time steps $\Delta t$, which are fractions of $\Delta z/c$. When the electron beam enters the plasma, it quickly comes to a radial equilibrium \cite{PoP24-023119}, and then its shape and its wakefield change slowly (figure~\ref{fig2-initial}). Once this happens, the fields can be updated less frequently. We merge the equilibrium electron beam and the fresh proton beam at this point and then propagate the two beams with their individual time steps in the plasma updated every $\Delta z = 200 c/\omega_p$. Since the period of betatron oscillations is proportional to the square root of the total particle energy, the time steps scale similarly and are $\Delta z / (2 c)$ for protons and $\Delta z / (256 c)$ for 18\,MeV electrons. In the presented runs, we initiate the electron beam at $z = -1000 c/\omega_p$ and follow its initial evolution up to $z = 1000 c/\omega_p$ with $\Delta z = 5 c/\omega_p$ or shorter.

\begin{figure}[tb]
\centering\includegraphics{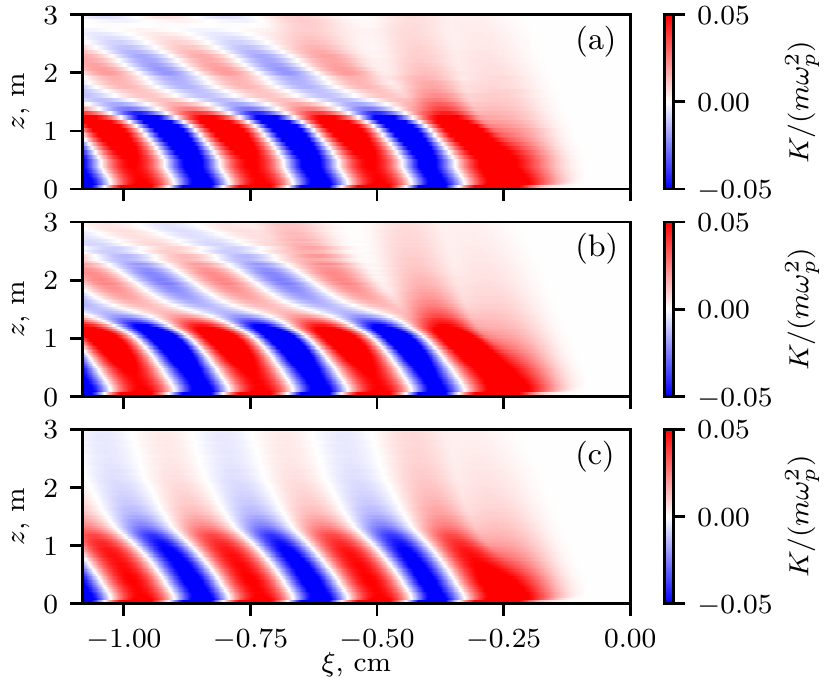}
\caption{The focusing strength $K (\xi, z)$ for $W_e = 18$\,MeV and $n_0 = 2 \times 10^{14}\text{cm}^{-3}$ calculated with: (a) $\Delta z = 200 c/\omega_p$ and beam substepping, (b) $\Delta z = 2 c/\omega_p$ and beam substepping (reference run), and (c) $\Delta z = c \Delta t = 2 c/\omega_p$ and no substepping. Initial beam evolution (up to $z = 1000 c/\omega_p \approx 37$\,cm) in all cases is simulated with $\Delta z = 0.5 c/\omega_p$ and beam substepping.}\label{fig3-substepping}
\end{figure}

While natural, the trick with beam substepping needs to be checked for correctness. Since the main purpose of the seed bunch is to focus or defocus proton beam slices, we compare the focusing strength $K =  F_\perp (r_f) / r_f$ at $r_f = 0.5 c/\omega_p$ in runs with different propagation steps for the lowest energy electron beam (figure~\ref{fig3-substepping}). Regular (a) and low-step (b) runs produce close results, while without substepping (c), even the step as short as $1/40$ of the betatron period in the cross-section of strongest focusing is insufficient.

\begin{figure}[tb]
\centering\includegraphics{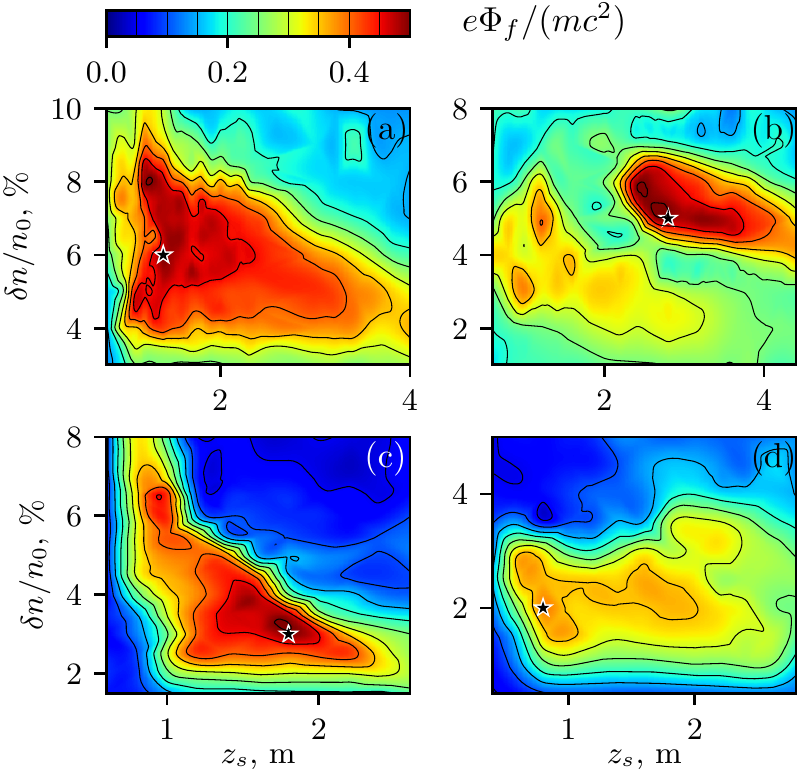}
\caption{Maps of the steady-state wakefield amplitude $\Phi_f (z_s, \delta n)$ for $W_e = 18$\,MeV (left), $W_e = 160$\,MeV (right), $n_0 = 2 \times 10^{14}\text{cm}^{-3}$ (top), and $n_0 = 7 \times 10^{14}\text{cm}^{-3}$ (bottom). Asterisks indicate the locations of the best variants, which are detailed in other figures.}\label{fig4-grid}
\end{figure}

To find the optimum density profile, we first locate the maximum of $\Phi_f$ in the three-dimensional parameter space $(z_p, L_s, \delta n)$ making large steps $\Delta z_p = \Delta L_s = 40$\,cm, $\delta n = 0.01 n_0$ on a rectangular grid, then reduce the steps and specify the maximum location, and finally scan the vicinity of the maximum to understand how wide the optimum area is. The result is most sensitive to height $\delta n$ and average position $z_s = z_p + L_s/2$ of the step (figure~\ref{fig1-geometry}), so we visualize the result as a surface $\Phi_f (z_s, \delta n)$, in which each point is maximized in $L_s$ (figure~\ref{fig4-grid}).

Even with the substepping tricks, simulation of one variant at low (high) plasma density takes about 80 (600) core-hours on Intel Xeon E5  processors. The total computational cost of this study is about $10^6$ core-hours.

\section{Results}
\label{s3}

As we see from figure~\ref{fig4-grid}, in three out of four cases it is possible to freeze the wakefield amplitude at the level of half the wavebreaking field. This corresponds to the limit imposed by the nonlinear elongation of the wave period \cite{PoP20-083119}: the surfaces $\Phi_f (z_s, \delta n)$ are as if cut off at this level. The lower-energy seed beam creates a strong field in wider parameter ranges, and the reason for this is discussed in section~\ref{s4}.

The optimum density increase $\delta n$ is higher than for SSM seeded by a sharp beam cut or ionization front. For the latter, the optimum is at $\delta n/n_0 \approx 2 / N$, where $N$ is the number of macro-bunches in the proton beam \cite{PoP22-103110}. In our case, taking $2 \sigma_{zb}$ as the beam length and putting $N = \omega_p \sigma_{zb} / (\pi c)$, we obtain $\delta n/n_0 \approx 0.03 \ (0.02)$ for the low (high) plasma density. This is almost twice smaller than the optimum values in all regimes (figure~\ref{fig4-grid}) except for the 160\,MeV beam in the high-density plasma [figure~\ref{fig4-grid}(d)].

\begin{figure}[tb]
\centering\includegraphics{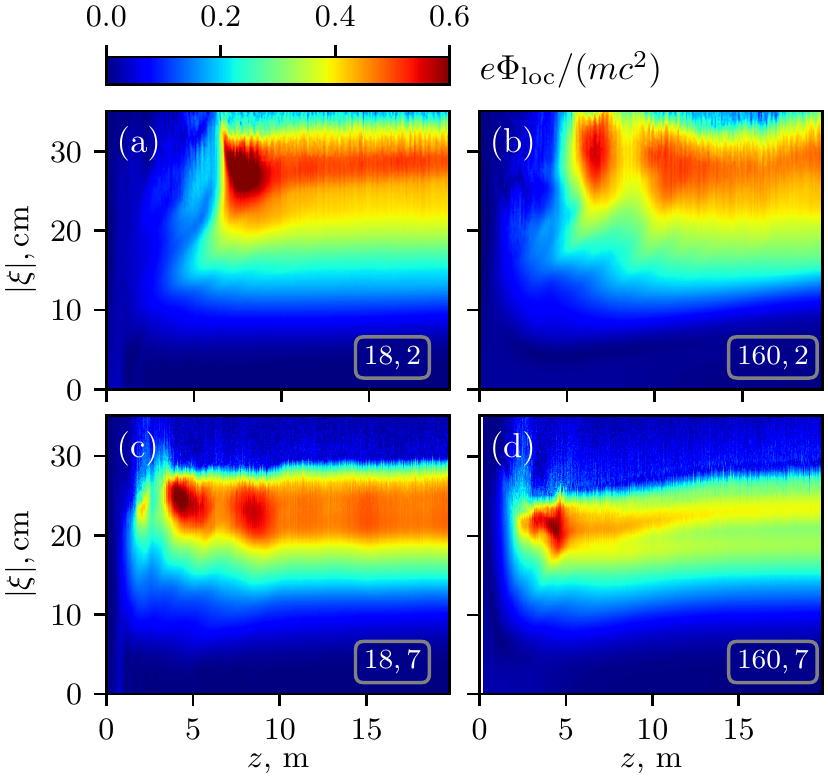}
\caption{Wakefield amplitude $\Phi_\text{loc} (z, \xi)$ for the four best variants. The legends show the electron beam energy $W_e$ (in MeV) and the plasma density $n_0$ (in $10^{14}\text{cm}^{-3}$).}\label{fig5-maps}
\end{figure}

\begin{figure}[tb]
\centering\includegraphics{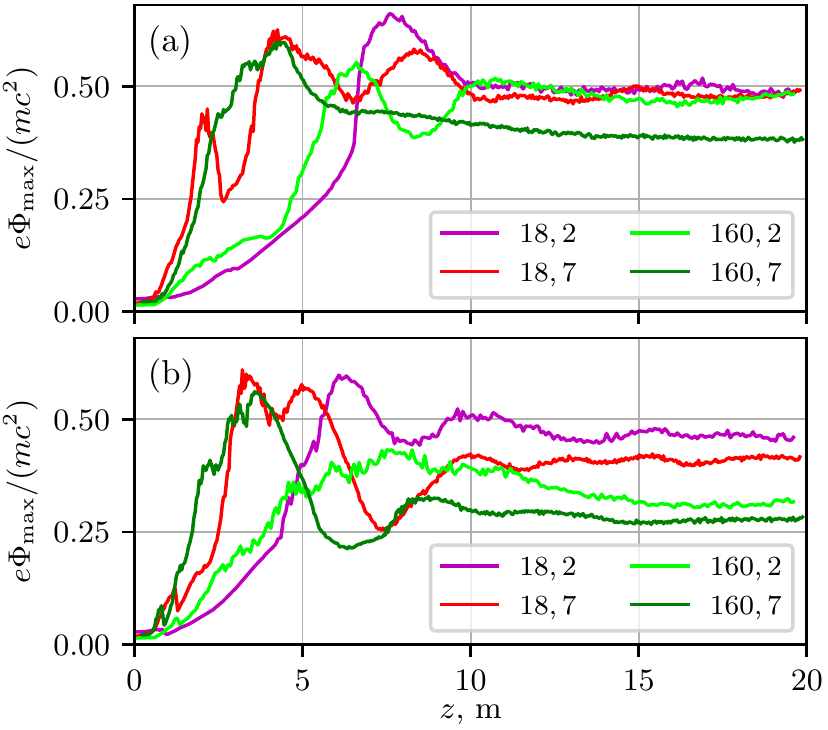}
\caption{Maximum wakefield amplitude $\Phi_\text{max} (z)$ for the best variants with smooth (a) and sharp (b) density profiles. The legends show the electron beam energy $W_e$ (in MeV) and the plasma density $n_0$ (in $10^{14}\text{cm}^{-3}$). }\label{fig6-gmaxf}
\end{figure}

\begin{figure}[tb]
\centering\includegraphics{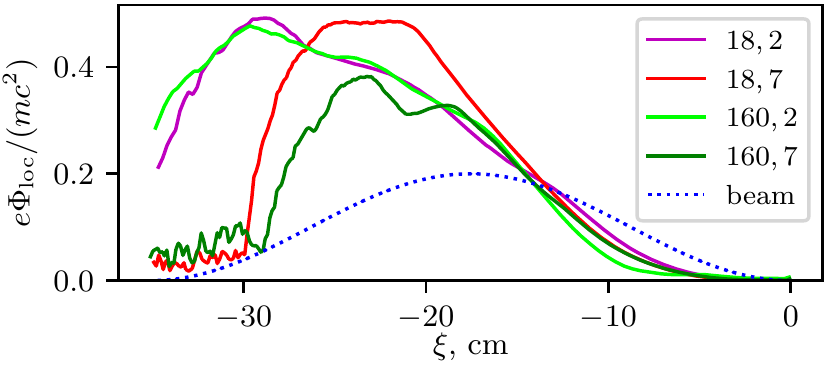}
\caption{Growth of the steady-state wakefield along the beam, $\Phi_\text{loc} (\xi)$ at $z = L_\text{max}$, for the best variants. The legend shows the electron beam energy $W_e$ (in MeV) and the plasma density $n_0$ (in $10^{14}\text{cm}^{-3}$). The blue dotted line shows the initial proton beam density (in arbitrary vertical units).}\label{fig7-lines}
\end{figure}

\begin{table*}[tb]
\caption{Parameters of the best variants with smooth (1st group) and steep (2nd group) density profiles.}\label{t2}
 \begin{center}\begin{tabular}{llllllll}\hline
  $W_e$, MeV & $n_0$, $10^{14}\text{cm}^{-3}$ & $\delta n / n_0$, \% & $z_p$, m & $z_s$, m & $L_s$, m & $e \Phi_f / (m c^2)$ & $E_f$, GV/m \\ \hline
18 & 2 & 6.0 & 0.8 & 1.4 & 1.2 & 0.49 & 0.68 \\
160 & 2 & 5.0 & 1.2 & 2.8 & 3.2 & 0.48 & 0.67 \\
18 & 7 & 3.0 & 0.8 & 1.8 & 2.0 & 0.49 & 1.27 \\
160 & 7 & 2.0 & 0.2 & 0.8 & 1.2 & 0.38 & 0.98 \\ \hline
18 & 2 & 5.5 & 1.2 & 1.2 & 0.0 & 0.46 & 0.64 \\
160 & 2 & 2.5 & 1.6 & 1.6 & 0.0 & 0.32 & 0.44 \\
18 & 7 & 3.0 & 1.4 & 1.4 & 0.0 & 0.42 & 1.07 \\
160 & 7 & 1.5 & 1.0 & 1.0 & 0.0 & 0.28 & 0.72 \\
  \hline
 \end{tabular}\end{center}
\end{table*}

\begin{figure}[tb]
\centering\includegraphics{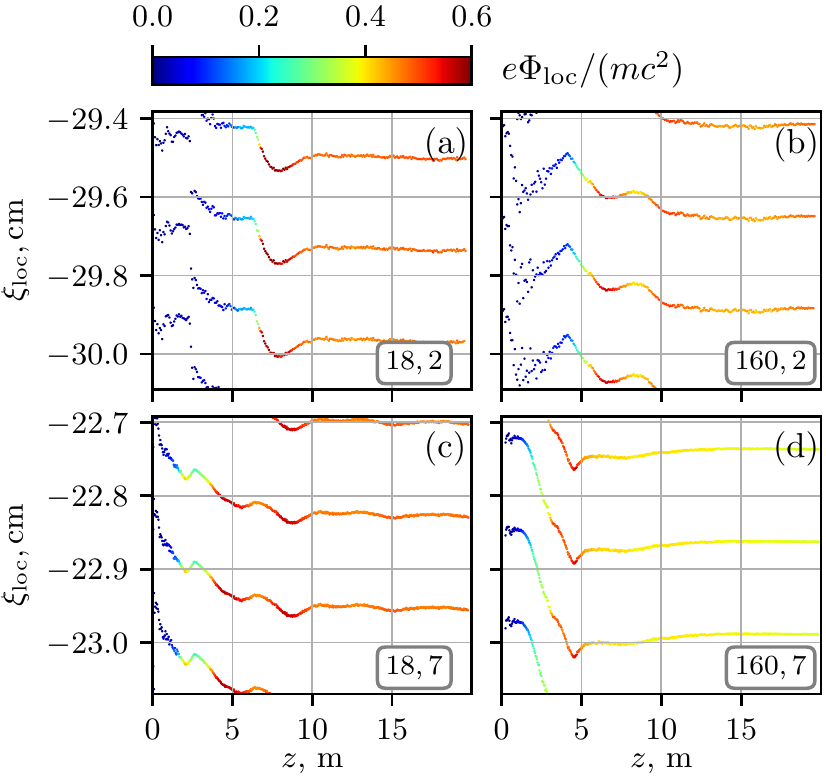}
\caption{Location of potential extrema $\xi_\text{loc} (z)$ in the regions of the strongest steady-state wakefield for the best variants. The color of points show the wakefield amplitude $\Phi_\text{loc}$. The legends show the electron beam energy $W_e$ (in MeV) and the plasma density $n_0$ (in $10^{14}\text{cm}^{-3}$).}\label{fig8-phase}
\end{figure}

\begin{figure}[tb]
\centering\includegraphics{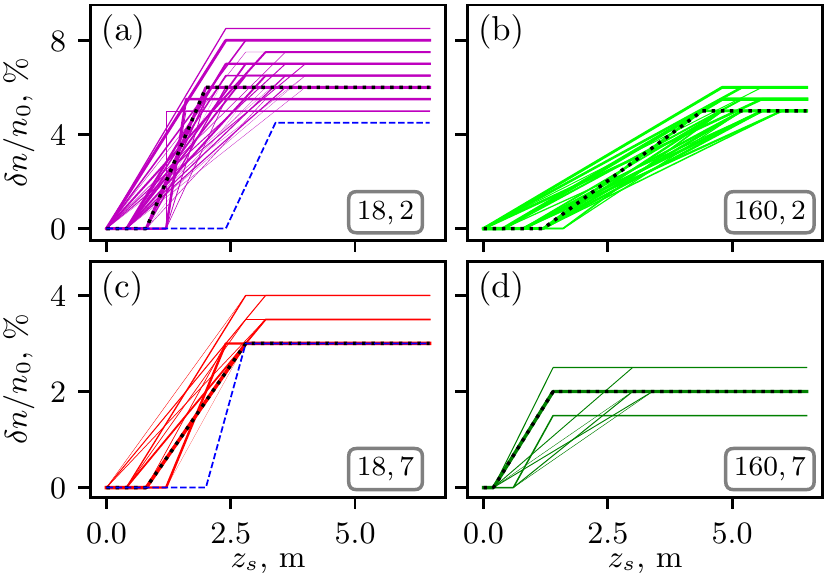}
\caption{Plasma density profiles that give a steady-state wakefield $\Phi_f$ above 90\% of the best variant values $\Phi_b$. The line width is proportional to $\Phi_f - 0.9\Phi_b$. Black dotted lines are density profiles for the best variants. The legends show the electron beam energy $W_e$ (in MeV) and the plasma density $n_0$ (in $10^{14}\text{cm}^{-3}$). Blue dashed lines in (a) and (c) are the best profiles for unseeded self-modulation.}\label{fig9-profiles}
\end{figure}

The best cases (marked by asterisks in figure~\ref{fig4-grid}) are detailed in figures~\ref{fig5-maps}--\ref{fig9-profiles} and table~\ref{t2}. Wakefield growth along the plasma is typical for low-emittance beams \cite{PoP22-123107}: the field first peaks at $z \sim 4$ (8)\,m (depending on the plasma density), then slightly decreases and remains approximately constant after $z = 10$\,m [figures~\ref{fig5-maps} and~\ref{fig6-gmaxf}(a)]. The steady-state wakefield grows almost linearly along the beam (figure~\ref{fig7-lines}). This means that most of the beam is self-modulated, and contributions of micro-bunches coherently add up. Having reached its maximum in $\xi$, the wakefield quickly disappears (figures~\ref{fig5-maps} and~\ref{fig7-lines}), because ion motion causes quick wavebreaking \cite{PRL109-145005,PoP21-056705}. The steady-state longitudinal electric field $E_f$ at the maximum depends on the plasma density and is of GV/m scale (table~\ref{t2}). Once the proton beam self-modulates, the phase velocity of the wakefield approaches $c$, but the wave is not perfectly phase-stable (figure~\ref{fig8-phase}).

The plateau-like character of the optimum in figure~\ref{fig4-grid} leaves a lot of freedom in choosing density profiles (figure~\ref{fig9-profiles}). However, all good profiles have a smooth density increase. If we restrict the search to sharp density steps only ($L_s=0$), then the wakefield in the best cases is noticeably weaker [figure~\ref{fig6-gmaxf}(b), table~\ref{t2}]. This contrasts with the ionization front seeding, for which the result is insensitive to the length of density transition \cite{PoP22-103110}.

\begin{figure}[tb]
\centering\includegraphics{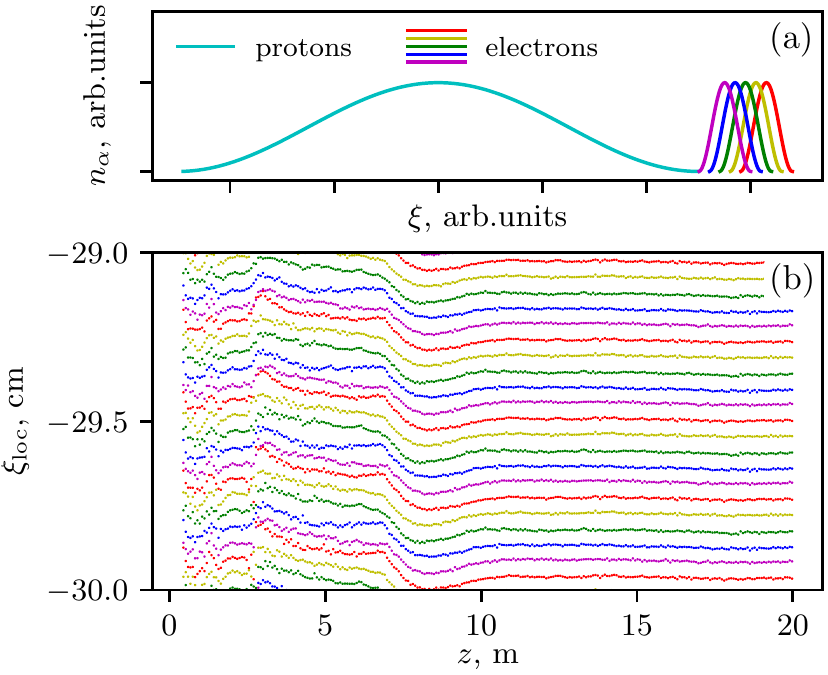}
\caption{Demonstration of phase-locking: (a) the idea of the test (the initial beam densities and sizes are not in scale), (b) location of potential extrema $\xi_\text{loc} (z)$ in the region of the strongest wakefield for these electron beam positions. The color code is the same in both fragments.  }\label{fig10-phshift}
\end{figure}

The steady-state wakefield of a self-modulated proton beam must be phase-locked to the seed electron bunch. We checked this for the 18\,MeV electron beam in the low-density plasma. Changing the distance between the beams leads to a corresponding shift of the wave relative to the proton beam (figure~\ref{fig10-phshift}). Uncontrolled self-modulation, if any, grows from the shot noise, which depends on the number of macro-particles in the proton beam. Real beams have much lower shot noise than simulated beams of large macro-particles \cite{PRST-AB16-041301}, so their self-modulation is even easier to control.

\begin{figure}[tb]
\centering\includegraphics{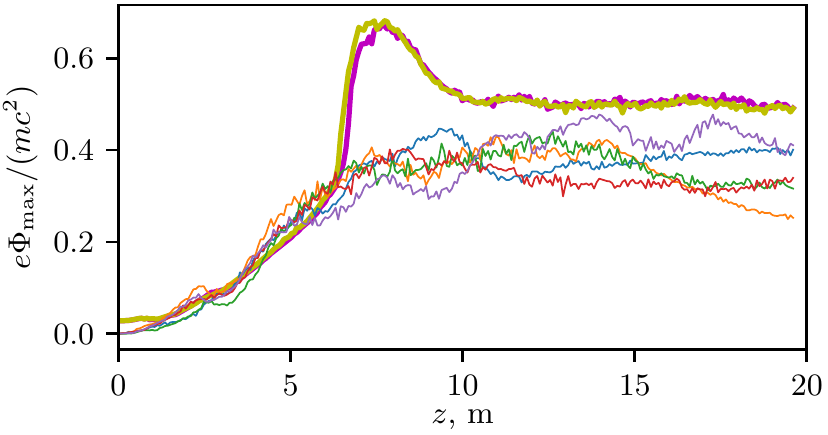}
\caption{Maximum wakefield amplitude $\Phi_\text{max} (z)$ for unseeded (thin lines) and seeded with the 18\,MeV electron beam (thick lines) self-modulation in optimally profiled plasmas of the initial density $n_0 = 2 \times 10^{14}\text{cm}^{-3}$. Variants differ in the initial values of the random number generator used to create the proton beam.}\label{fig11-noseed}
\end{figure}

If there is no seed perturbation, the proton beam self-modulates anyway, and profiling of the plasma density also increases the wakefield at large propagation distances. In this case, the optimum increase in density begins later [figure~\ref{fig9-profiles}(a),(c)], since it takes longer for the initial self-modulation to grow from the noise level. However, the result depends on this noise, and proton beams created with different initial values of the random number generator produce different wakefields (figure~\ref{fig11-noseed}).

\begin{figure}[tb]
\centering\includegraphics{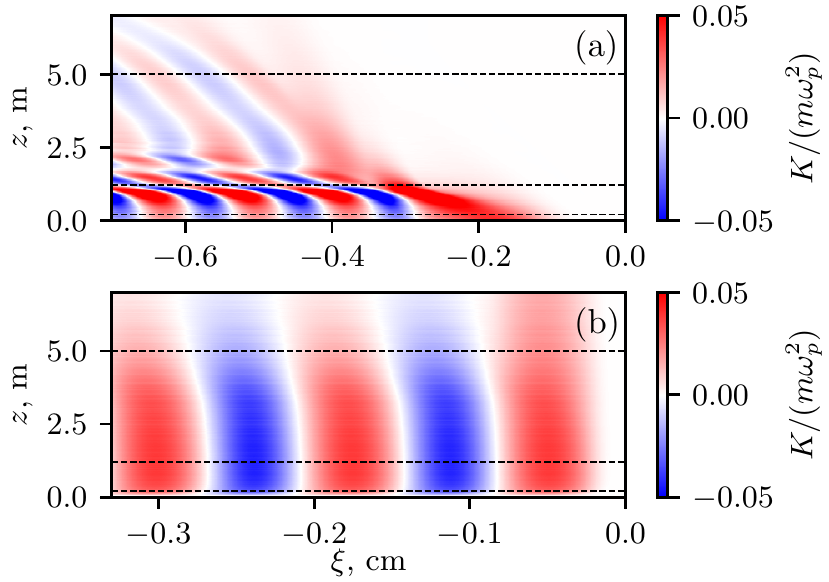}
\caption{The focusing strength $K (\xi, z)$ for (a) $W_e = 18$\,MeV and (b) $W_e = 160$\,MeV electron beams in $n_0 = 7 \times 10^{14}\text{cm}^{-3}$ plasma. Dashed horizontal lines show propagation distances $z$ detailed in figure~\ref{fig13-eevol}.}\label{fig12-kevol}
\end{figure}

\begin{figure}[tb]
\centering\includegraphics{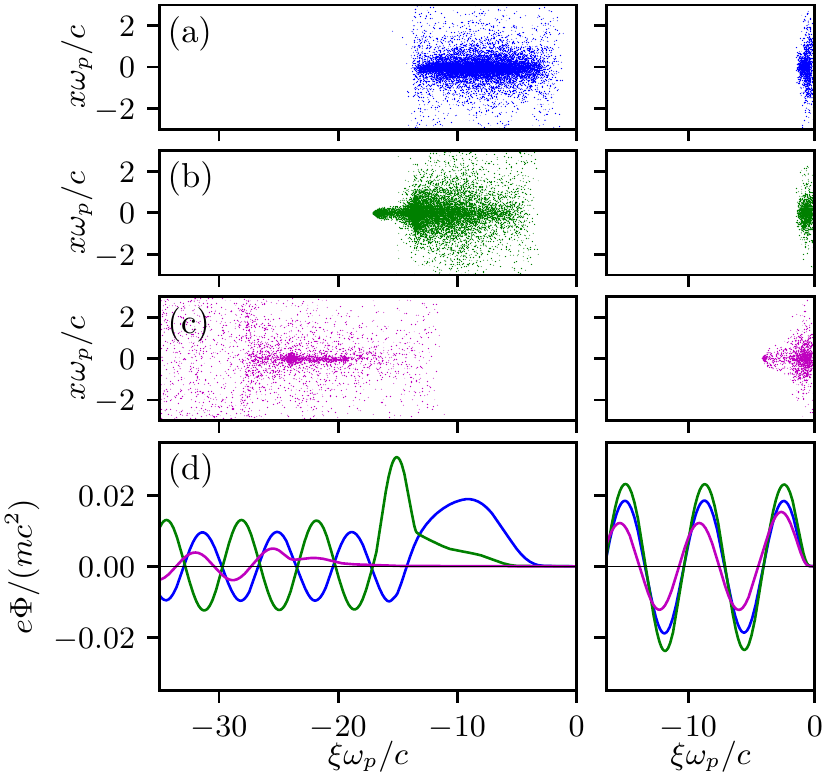}
\caption{Portraits of 18\,MeV (left) and 160\,MeV (right) electron beams at (a) $z=0.2$\,m, (b) $z=1.2$\,m, and (c) $z=5$\,m in the high-density plasma ($n_0 = 7 \times 10^{14}\text{cm}^{-3}$); (d) corresponding (of the same color) wakefield potential $\Phi (\xi)$ on the axis. }\label{fig13-eevol}
\end{figure}

\section{Discussion}
\label{s4}

Optimization of density profiles revealed an unexpected result. A high-quality electron beam with an energy of 160\,MeVs turns out to be a worse seed than a longer 18\,MeV beam. For the 160\,MeV beam, the strongest possible wakefield is reached in a smaller region of the parameter space for the low plasma density, and not reached at all for the high plasma density (figure~\ref{fig4-grid}). The reason lies in the lifetime of the seed wakefield: for the 160\,MeV beam it lasts longer (figure~\ref{fig12-kevol}). The 18\,MeV beam quickly loses energy and disappears, while the 160\,MeV beam produces a strong seed wave of an unchanged phase even after partial destruction (figure~\ref{fig13-eevol}).

\begin{figure*}[tb]
\centering\includegraphics{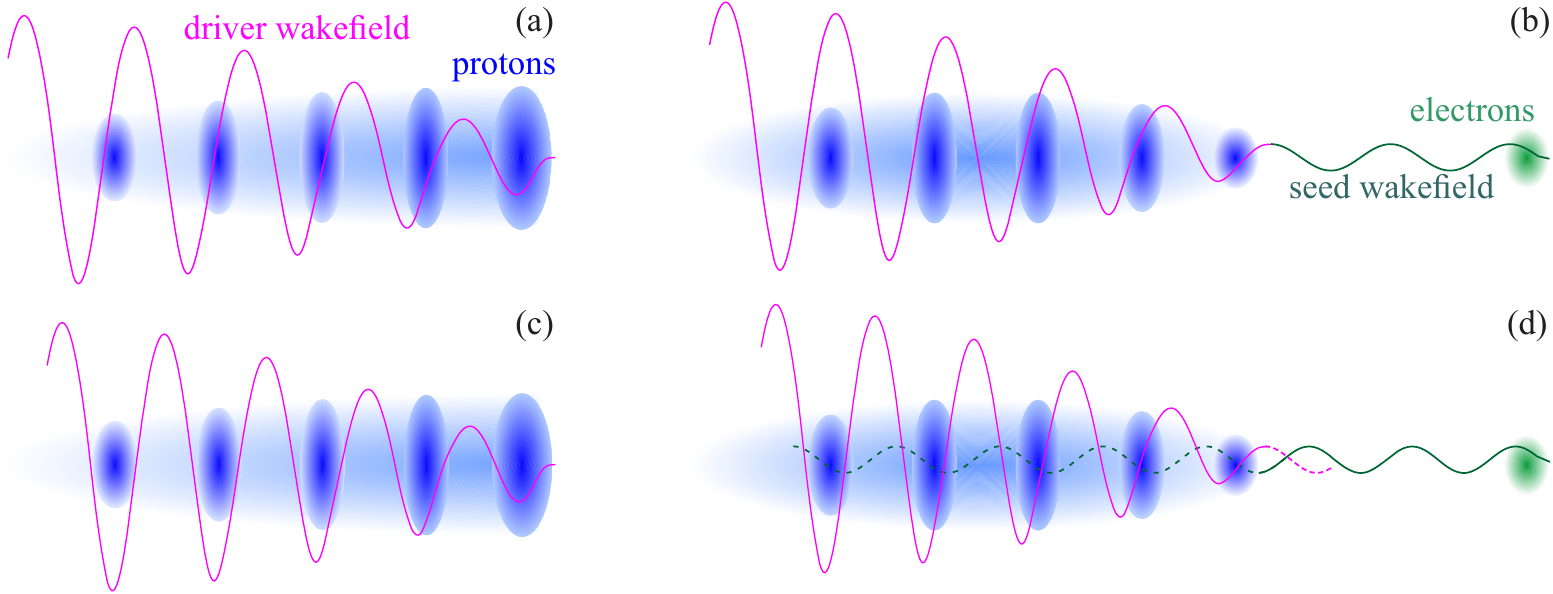}
\caption{Differences between leading edge seeding (a), (c) and electron beam seeding (b), (d). The wakefield potential before (a), (b) and after (c), (d) the density increase is shown schematically in relation to growing micro-bunches.}\label{fig14-idea}
\end{figure*}

To understand how a long lifetime of the seed wave can worsen the self-modulation, we compare this process in cases of leading edge seeding and electron beam seeding (figure~\ref{fig14-idea}). Consider the leading edge seeding first. Before the density increases, the micro-bunches begin to form in the decelerating phases of the growing wakefield. The latter, in turn, is phase locked to the first micro-bunch located near the leading edge of the proton beam [figure~\ref{fig14-idea}(a)]. After increasing the density, the wave remains phase locked to the first micro-bunch, but the wavelength is shortened, so that the favorable wave phase (focusing and decelerating) shifts forward to micro-bunch positions \cite{PoP22-103110} [figure~\ref{fig14-idea}(c)]. This mechanism works for any lifetime of the seed perturbation, because the seed wave produced by the leading edge and the wakefield of the first ``big'' micro-bunch are always in phase.

Now consider the electron beam seeding. Before the density increases, the micro-bunches are phase-locked to the seed wave produced by the electron bunch. Their own wakefield perfectly matches the seed wakefield, as it developed from the latter [figure~\ref{fig14-idea}(b)]. However, after increasing the density, the two wakefields become mismatched because of the large distance between the electron bunch and the body of the proton beam. The micro-bunches cannot quickly disappear in one place and appear in another, so their own wakefield remains almost the same as it was before the plasma density change. This wakefield is locked to the first ``big'' micro-bunch, and it becomes out of phase with the seed wakefield [figure~\ref{fig14-idea}(d)]. As a result, the wakefield of the electron bunch destroys the micro-bunches instead of helping them to grow. At some conditions, the micro-bunches ``win'', and the established wakefield is as high as the saturation level, but in a narrow region of the parameter space [figure~\ref{fig4-grid}(b)]. At others, the micro-bunches are destroyed stronger, and the wakefield is lower [figure~\ref{fig4-grid}(d)].

\begin{figure*}[tb]
\centering\includegraphics{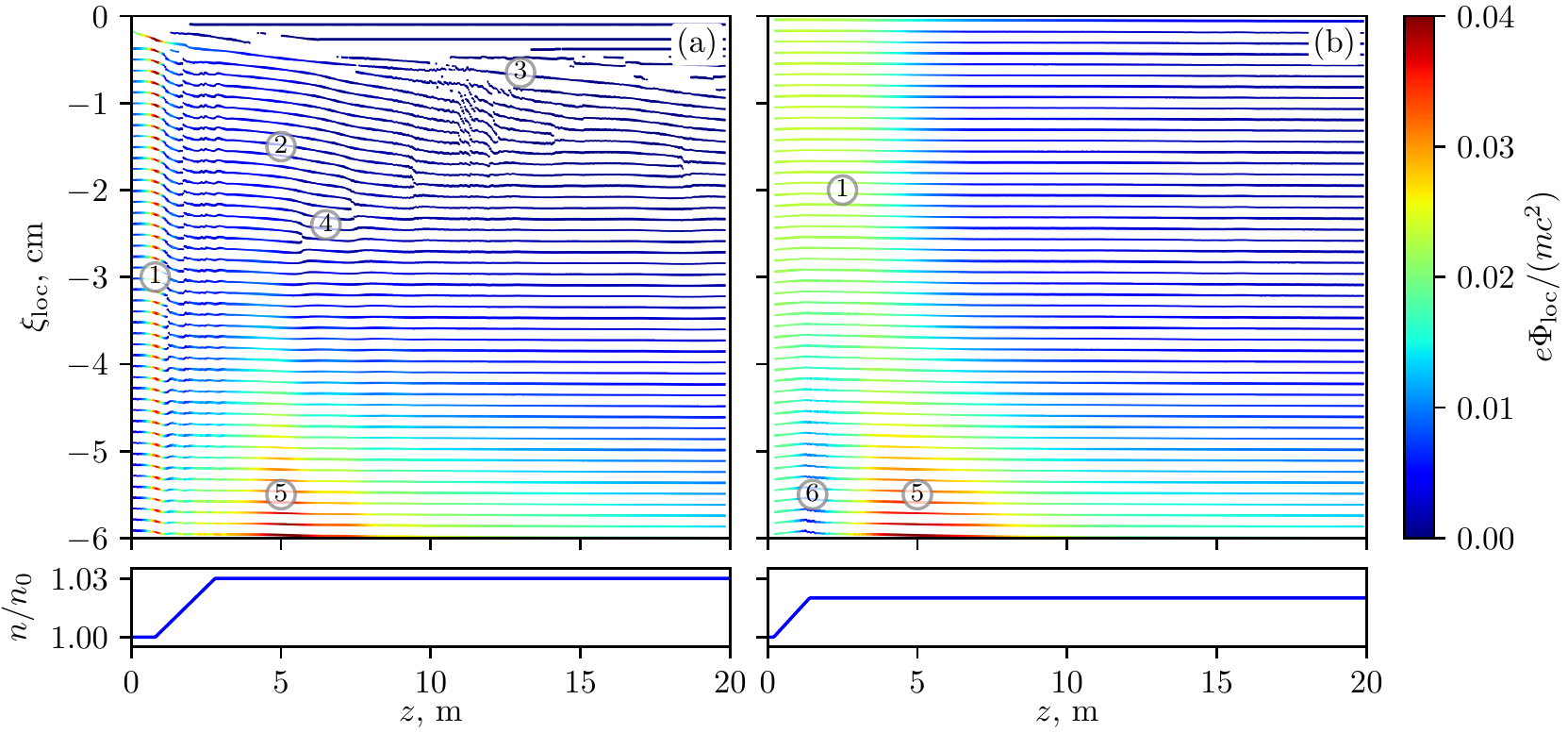}
\caption{Location of potential extrema $\xi_\text{loc} (z)$ and wakefield amplitude $\Phi_\text{loc}(z)$ at the beginning of the proton beam for best variants with $W_e = 18$\,MeV (a) and $W_e = 160$\,MeV (b) in the plasma with $n_0 = 7 \times 10^{14}\text{cm}^{-3}$. Bottom fragments show the plasma density profiles. The numbers in circles indicate the features discussed in the text.}\label{fig15-ssmmap}
\end{figure*}

We can observe the destructive interference of seed and micro-bunch wakefields, if any, on the maps of the wakefield potential (figure~\ref{fig15-ssmmap}). These maps conveniently show how wakefield phase and amplitude change along the beam and plasma. We first consider the case of 18\,MeV electrons [figure~\ref{fig15-ssmmap}(a)]. The seed wave is a narrow vertical color region in the figure (feature 1), which quickly grades, as $z$ increases, into a low-amplitude, low-phase velocity wakefield of the depleted electron beam (2). The line (3) corresponds to the weak wakefield of the very head of the electron beam, and its slope shows the velocity of 18\,MeV electrons that do not slow down. The seed wave delivers a quick push to protons, which initiates a growing modulation of the proton density. This modulation is phase-locked to the initial push. As the density modulation grow, its wakefield at some point becomes stronger than the wakefield of the depleted electron beam, and this transition appears in figure~\ref{fig15-ssmmap}(a) as quick changes of the wave phase (4). On further growth of beam modulation, the own wakefield of the micro-bunches starts to affect micro-bunch formation, and the exponential stage of self-modulation begins (5). The plasma density increase has no effect on the described initial stages of self-modulation, as the action of the seed wave ends before the density changes.

In the case of 160\,MeV electrons [figure~\ref{fig15-ssmmap}(b)], the seed wave lasts much longer (1) and passes into the exponentially growing wakefield (5) without large phase jumps. However, when the plasma density begins to increase, the wakefield amplitude goes down (feature 6) because of the destructive interference mentioned before. This feature obviously impedes the development of self-modulation. If we delay the density increase to $z_p \gtrsim 5$\,m, the self-modulation will develop to the stage of micro-bunch destruction in a uniform plasma \cite{PoP22-103110}, and the steady-state wakefield will still be low.

As follows from the above considerations, the optimum density profiles in the case of 160\,MeV electron beam are not universal and depend on the distance $\xi_e$ between the beams. This contrasts with the 18\,MeV beam case, for which this distance is unimportant.

We see that high-energy electron beams work worse as seeds for SSM than low-energy ones, but where is the boundary between high and low energies? To find it, we compare the distances of proton beam micro-bunching and electron beam depletion. For the highest steady-state wakefield, an increase in density must begin at the stage of exponential field growth \cite{PoP22-103110}, that is, before
\begin{equation}\label{e4}
    L_b \sim \sqrt{\frac{W_b n_0}{mc^2 n_b}} \frac{c}{\omega_p}.
\end{equation}
If the electron beam length is matched to the plasma wavelength \cite{PoP12-063101} ($\sigma_{ze} \sim c/\omega_p$), and the beam is not too narrow ($\sigma_{re} \sim c/\omega_p$), the amplitude of the seed wave is $E_z \sim E_0 n_e / n_0$, and the depletion length is
\begin{equation}\label{e5}
    L_d \sim \frac{W_e}{e E_z} \sim \frac{W_e n_0}{m c^2 n_e} \frac{c}{\omega_p}.
\end{equation}
Thus, the boundary energy $W_0$, at which $L_b \sim L_d$, is the geometric average:
\begin{equation}\label{e6}
    \frac{W_0}{n_e} \sim \sqrt{\frac{W_b}{n_b} \frac{mc^2}{n_0}}.
\end{equation}
For the 18\,MeV beam and low (high) density plasma, we find $W_0 = 60$ (30) MeV, so the seeding is efficient. If the electron beam is shorter than $c/\omega_p$, then the decelerating field and the boundary energy $W_0$ are $\sigma_z \omega_p/c$ smaller \cite{PoP12-063101}. This is the case for our 160\,MeV beam, for which $W_0 = 30$\,MeV, regardless of the plasma density. The wakefield of this beam lasts longer than necessary for efficient seeding.

\section{Summary}
\label{s5}

We numerically optimized the longitudinal profile of the plasma density in search of the strongest wakefield established after proton beam self-modulation seeded by a short electron beam. To speed up simulations, we used a quasi-static code, long steps for calculating the plasma response to the beams, and substepping for propagating the electron beam.

For beam parameters discussed in the context of future AWAKE experiments, it is possible to ``freeze'' the wakefield at approximately half the wavebreaking level. The self-modulation occurs during the first 10\,m of beam propagation in the plasma, after which the beam excites a wakefield with a constant phase velocity (close to the speed of light) and a constant amplitude. This steady-state wakefield is phase-locked to the seed electron bunch. In all studied regimes, the optimum plasma profiles are those with a smooth density increase at the early times of proton beam micro-bunching.

The self-modulation is more efficient, if the seed beam is depleted before the density increase. Otherwise, the wakefield of the electron beam destructively interferes with the growing wakefield of the micro-bunches and either reduces the variety of acceptable plasma profiles or reduces the achievable steady-state amplitude of the wakefield.

\ack

This work was supported by the Russian Science Foundation, project 20-12-00062. Simulations were performed on HPC-cluster "Akademik V.M. Matrosov" \cite{matrosov}.

\end{document}